\documentclass[pra,twocolumn,floatfix,superscriptaddress,eqsecnum]{revtex4-1}

\usepackage{graphicx}
\usepackage{color}
\usepackage{amsmath,amssymb,amstext} 

\usepackage[caption=false]{subfig}

\newcommand{\degree}{\ensuremath{^\circ}}
\newcommand{\dg}{^\dagger}
\newcommand{\ket}[1]{|#1\rangle}
\newcommand{\bra}[1]{\langle#1|}

\begin{document}

\title{Tunable coupling between three qubits as a building block for a superconducting quantum computer}

\author{Peter Groszkowski}
\affiliation{Institute for Quantum Computing and Department of Physics and Astronomy, University of Waterloo, Waterloo, Ontario, Canada N2L 3G1}

\author{Austin G. Fowler}
\affiliation{Centre for Quantum Computation and Communication Technology, School of Physics, The University of Melbourne, Victoria 3010, Australia}

\author{Felix Motzoi}
\affiliation{Institute for Quantum Computing and Department of Physics and Astronomy, University of Waterloo, Waterloo, Ontario, Canada N2L 3G1}

\author{Frank K. Wilhelm}
\affiliation{Institute for Quantum Computing and Department of Physics and Astronomy, University of Waterloo, Waterloo, Ontario, Canada N2L 3G1}


\begin{abstract}
Large scale quantum computers will consist of many
interacting qubits. In this paper we expand the two flux qubit
coupling scheme first devised in [Phys. Rev. B {\bf 70}, 140501 (2004)] and realized
in [Science {\bf 314}, 1427 (2006)]  to a three-qubit, two-coupler scenario. We study
L-shaped and line-shaped coupler geometries, and show how the
interaction strength between qubits changes in terms of the
couplers' dimensions. We explore two cases: the ``on-state" where
the interaction energy between two nearest-neighbor qubits is
high, and the ``off-state" where it is turned off. In both
situations we study the undesirable crosstalk with the third
qubit. Finally, we use the GRAPE algorithm to find efficient pulse
sequences for two-qubit gates subject to our calculated physical constraints on
the coupling strength.
\end{abstract}

\maketitle

\section{Introduction}
\label{sec:Introduction}

Quantum computers hold great promise to dramatically alter the way we
perform computation. They can be constructed from qubits, two-level
quantum systems which act as their fundamental building blocks. Although there are different approaches in actually building two-level quantum systems, in recent years superconducting devices have seen some considerable success. There have been multiple
proposals for designs of superconducting qubits and of their possible coupling mechanisms {\cite{Orlando99,
  Nakamura99,Makhlin01,Vion02, Chiorescu03, Martinis02, Clarke2008,
  Devoret04, wendin-2005, You2006}. In this paper, we look at inductively
coupled flux qubits along with dc Superconducting QUantum Interference
Devices (DC-SQUIDs) used for tunability. The DC-SQUIDs are treated as
high excitation energy quantum objects and are assumed to always stay
in their ground states \cite{Ashhab2008,hutter2006tcq}. This method
was first theoretically described in \cite{Plourde2004}, and
subsequently demonstrated experimentally in \cite{Hime2006}. Similar coupling
approaches were presented in {\cite{Niskanen07,harris2007,Ploeg2007,brink2005,Peropadre2010}.

This paper is organized as follows. Section \ref{sec:Setup} describes the system under study, in particular the different geometries that are explored. Section \ref{Theory} looks at the mathematics of coupling and describes how it varies in terms of physical system parameters. Next, Section \ref{CouplingResults} shows numerical results when specific experimentally achievable parameters are used. Finally Section \ref{QubitControl} looks at the application of the GRAPE algorithm when performing two-qubit gates. The conclusion is presented in Section \ref{Conclusion}.

\section{Setup}
\label{sec:Setup}

A flux qubit can be constructed from a superconducting loop interrupted by three Josephson junctions \cite{Orlando99}. A current flowing in one direction can represent a computational basis quantum state $\ket{0}$ and in the opposite direction, a state $\ket{1}$. The strength of the coupling between qubits is mediated by a circulating current in a DC-SQUID which is inductively coupled to each of the qubits. This circulating current can be controlled by the applied flux as well as the bias current, both of which can be tuned experimentally. Furthermore, the DC-SQUIDs may also be used for qubit readout \cite{PhysicaC02,Lupascu06,vanderWal00}.

\begin{figure}
    \begin{center}
  \subfloat[]{\includegraphics[width=7cm]{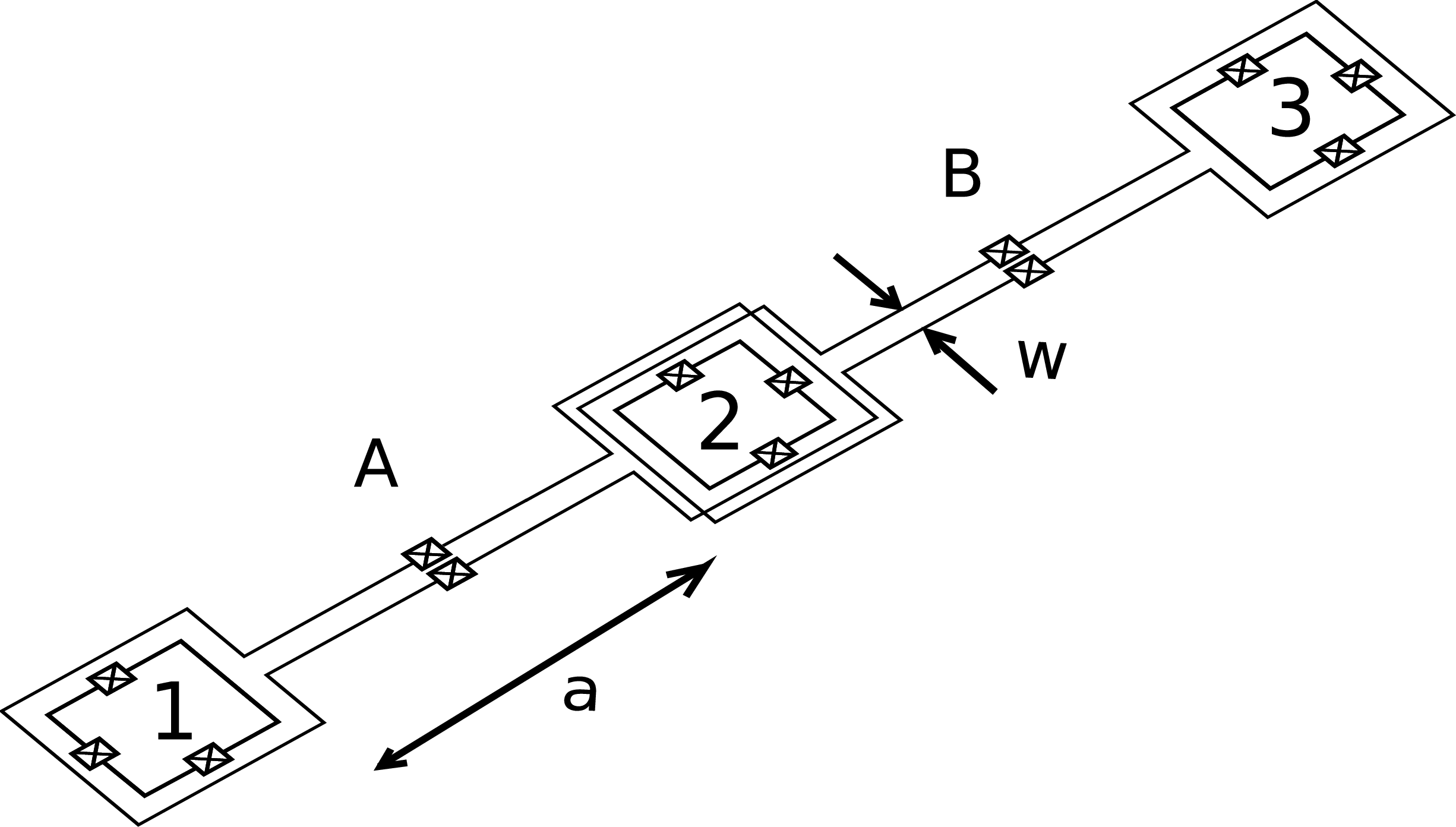}} 
  \vspace{0.2cm}
  \subfloat[]{\includegraphics[width=7cm]{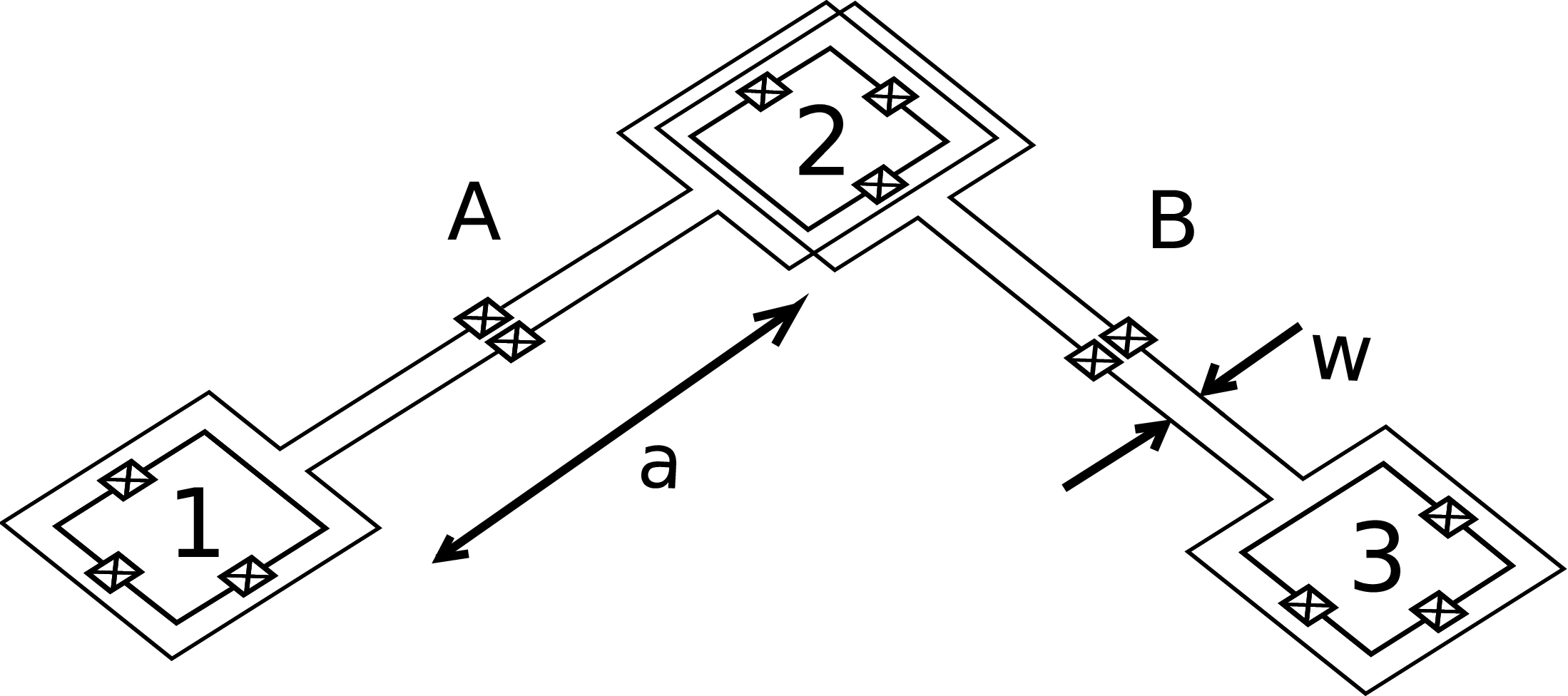}}                
  \vspace{0.2cm}
  \end{center}
  \caption{Two different arrangements of flux qubits 1, 2, 3 and couplers A, B. ``$a$'' represents the arm length of the couplers, and ``$w$'' their width. We assume that the wires of couplers A and B overlap at most in two places around qubit 2. a) Straight-line shape. b) Capital ``L'' shape. }
    \label{fig:CouplerGeometry}
\end{figure}

The performance of any inductively coupled qubit architecture will
strongly depend on the geometry of the qubit-coupler layout. The
couplers themselves are based on the long-distance design proposed in \cite{Fowler07}. In this paper we study two geometries, which can potentially serve as building blocks for topologically motivated approaches to quantum computation \cite{Kitaev1995,Raussendorf2007b} with high error thresholds \cite{Dennis2002, Raussendorf2007b,Fowler2009-1,Wang2010}. In these approaches the physical qubits are typically laid out in a (for example) 2D square lattice. The geometries we study, which may be used to build such a lattice, are both shown in Fig.~\ref{fig:CouplerGeometry}. In a), the coupling DC-SQUIDs form a straight line and in b), they enclose a 90\degree \ angle reminiscent of a capital letter ``L''. In both cases we have two qubits at the end points (labeled ``1'' and ``3'') and the third (labeled ``2'') in the middle. The middle qubit is therefore surrounded by both of the couplers. We consider the qubits at the edge to be nearest-neighbours to the middle qubit. The qubits are placed inside the couplers, hence maximizing the qubit-coupler inductance. We take the edge length of the square qubits to be $44\mu m$, roughly the same as those used in \cite{Hime2006}. The distance between the qubit and the surrounding wire of the coupler is between $2$ and $4 \mu$m. To understand what impact qubit separation has on the interaction energy, we vary both the couplers' arm length and arm width (shown as ``$a$'' and ``$w$'' respectively in Fig.~\ref{fig:CouplerGeometry}). The arm length $a$ varies between $25\mu$m and $300 \mu$m and we study cases of the arm width $w=48\mu$m and  $w=2 \mu$m (although in the latter case we keep a $10 \times 10 \mu m ^{2}$ loop in the center of the arm to simplify delivering flux through the couplers. We stress that even though the coupling DC-SQUIDs are the same in both geometries, we expect the coupling energies to differ as even for the same length couplers, the distance between qubit 1 and 3 will be smaller in the 'L' shaped scenario. In order to numerically calculate the inductance for any given configuration, we perform finite element simulations using FastHenry \cite{Kamon1993fasthenry} and assume we have aluminum wires with a penetration depth of $51$nm \cite{Biondi1959}. The result is a $5 \times 5$ inductance matrix (see section \ref{CouplingResults} for an example). Since qubit $2$ is always surrounded by two DC-SQUIDs, an experimental realization of this scenario would require crossing the wires of the couplers in at least two places. For simplicity we assume a fully symmetric situation meaning the qubits that are situated at the edges look the same to the qubit in the middle.

\begin{figure}[ht]
    \includegraphics[width=8cm]{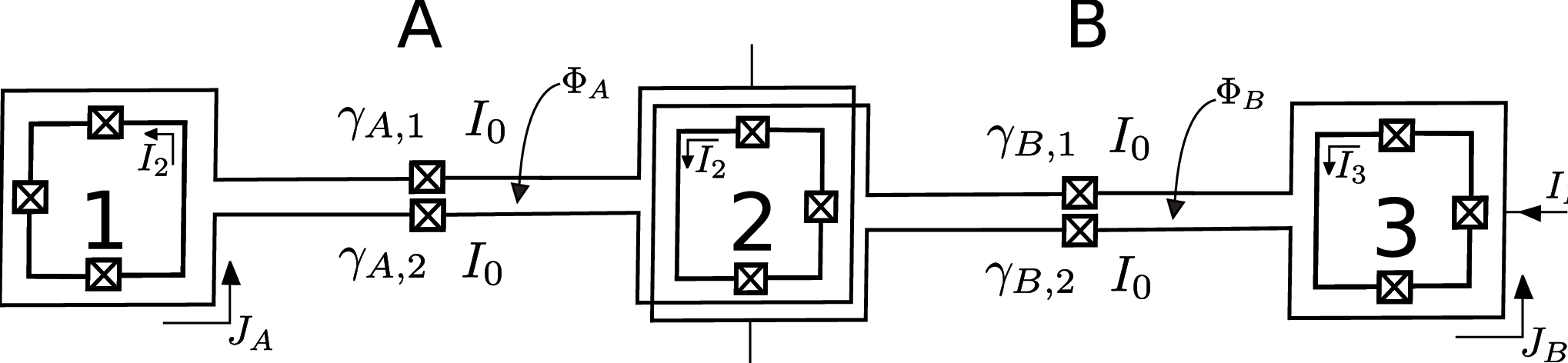}
    \caption{The three flux qubits labeled 1, 2 and 3, and two DC-SQUIDs labeled A and B, form a three-qubit, two-coupler system. $\gamma_{n,i}$ is the phase difference across the $i$th junction in a coupler $n$. $\Phi_{n}$ and $J_n$ are the applied flux and circulating current of coupler $n$ respectively. $I_{0}$ is the critical current of the Josephson junctions used in the coupling DC-SQUIDs --- for simplicity we assume that all the junctions, and hence their critical currents, are the same.  }
    \label{fig:ThreeQubitCouplerLineDetails}
\end{figure}

\section{Theory}
\label{Theory}
In this section we expand on the results presented in
\cite{Plourde2004}. First, we define the parameters of our system of
which a schematic representation is shown in
Fig.~\ref{fig:ThreeQubitCouplerLineDetails}. Each variable 
$\gamma_{n,i}$ is the phase difference across the $i$th junction in coupler
$n$. $\Phi_{n}$ is the applied flux threaded through coupler $n$,
$I_n$ its bias current and $J_n$ its circulating current. $I_{0}$ is
the critical current of the Josephson junctions used in the coupling
DC-SQUIDs --- for simplicity we assume that all the junctions, and
hence their critical currents, are the same. We further assume that we
are able to control the flux independently through each of the
DC-SQUIDs and qubits. This may need multiple bias coils and careful calibration to cancel
cross-control effects. We can write an effective Hamiltonian which describes the $j$th flux qubit \cite{Orlando99} as
\begin{equation}
\label{QubitH}
H_j=\frac{1}{2}(\epsilon_j \sigma_{z} + \Delta_j \sigma_{x})
\end{equation}
where $\sigma_{z}$ and $\sigma_{x}$ are Pauli matrices, $\epsilon_j$ is the energy bias which depends on the flux threaded through the qubit and $\Delta_j$ is the tunneling energy, which we take to be fixed at fabrication time and usually of the order of a few GHz. We stress that modified flux qubit designs, with variable $\Delta_j$ have been proposed \cite{Orlando99} and demonstrated experimentally \cite{paauw2009tuning,fedorov2011tuned,poletto2009coherent,chiarello2005superconducting}, but in principle one only needs control of one of the qubit's quadratures to do arbitrary single-qubit gates. Next, we note that each qubit $i$ interacts with any other qubit $j$ via the direct mutual inductance as well as via the mutual inductance mediated by the couplers. We can therefore write the total Hamiltonian of the system as
\begin{equation}
H=\sum_{j=1,2,3} H_j - \sum_{i \ne j}  K_{ij} {{\sigma^{i}_z} \otimes {\sigma^{j}_z}} \label{TotalH}
\end{equation}
where $H_j$ are the Hamiltonians of the individual qubits, and $K_{ij}$ correspond to the coupling energy between qubits $i$ and $j$ \cite{Orlando99}. Each $K_{ij}$ can be further described as
\begin{equation}
\label{eq:Ks}
    K_{ij} = M_{ij} I_{i}I_{j} - \sum_{n=A,B} \sum_{k=A,B} M_{n,i} M_{k,j}  \frac{\partial J_{n}}{\partial \Phi_{k}} I_{i}I_{j}
\end{equation}
with $M$ the inductance matrix of the system. We easily see that the first term corresponds to the direct coupling between the qubits, and the other terms are involved in the indirect interaction through the coupling DC-SQUIDs. The negative sign in front of the second term is justified in \cite{Ferber2005,ferber2010efficient}. Each of the couplers $n$ (for $n={A,B}$) can be controlled by either applying a bias current $I_n$ and/or a bias flux $\Phi_n$. By adjusting these two parameters we will study how the coupling between any pair of qubits can be either enhanced or reduced. The interaction energy in Eq. \ref{eq:Ks} involves the transfer functions of the form $\frac{\partial J}{\partial \Phi}$ which we have to calculate.

We start with Kirchhoff's current laws for coupler $n$. In the slowly varying limit of the junctions \cite{Plourde2004}, we can write
\begin{equation}
\label{kir1}
\begin{array}{rcl}
I_{n} & = & I_{0} \sin \gamma_{n1}+I_{0} \sin\gamma_{n2} \\
2J_{n} & = & I_{0} \sin \gamma_{n1}-I_{0} \sin\gamma_{n2}.
\end{array}
\end{equation}
Next, using $ \gamma_{n-} =\frac{\gamma_{n1}- \gamma_{n2}}{2} $ and $ \gamma_{n+} =\frac{\gamma_{n1}+ \gamma_{n2}}{2} $, we can rewrite Eqs. \ref{kir1} as
\begin{equation}
\label{kir2}
\begin{array}{rcl}
I_{n} & = & 2I_{0} \sin \gamma_{n+} \cos \gamma_{n-} \\
\end{array}
\end{equation}
and 
\begin{equation}
\label{kir2b}
\begin{array}{rcl}
J_{n} & = & I_{0}\sin{\gamma_{n-}}\cos{\gamma_{n+}}.
\end{array}
\end{equation}

From flux quantization \cite{deGennes66,Tinkham96}, we have another condition, which relates the flux passing through the DC-SQUID to phase differences across the junctions. This lets us write
\begin{equation}
\label{constraint1} 
\begin{array}{rcl}
\gamma_{n-} & = & \frac{\pi}{\Phi_{0}}(\Phi_{n} -
M_{n}J_{n} - M_{nk}J_{k}). \\
\end{array}
\end{equation}
By fixing the applied flux $\Phi_n$ and bias current $I_n$ we can
solve Eqs. \ref{kir2}, \ref{kir2b} and \ref{constraint1} to obtain
circulating currents $J_n$ as well as the phases $\gamma_{n-}$ and
$\gamma_{n+}$. Next we can implicitly differentiate Eqs. \ref{kir2},
\ref{kir2b} and \ref{constraint1} in order to obtain the transfer
functions $\frac{\partial J_{n}}{\partial \Phi_{n}}$ and
$\frac{\partial J_{n}}{\partial \Phi_{k}}$ (here $k$ is just a coupler
index with $k\neq n$), which we will need to calculate the interaction energy $K_{ij}$. Noting that the bias current does not depend on the flux, and differentiating Eq. \ref{kir2} gives
\begin{equation}
\frac{\partial I_{n}}{\partial \Phi_{k}} = 0 =2I_{0} \frac{\partial}{\partial \Phi_{k}} \left( \sin{\gamma_{n+}}\cos{\gamma_{n-}} \right)
\end{equation} 
which further leads to
\begin{equation}
\label{eq:kir4}
\frac{\partial \gamma_{n+}}{\partial \Phi_{k}} = \tan^{2}\gamma_{n+} \tan^{2}\gamma_{n-} \frac{\partial \gamma_{n-}}{\partial \Phi_{k}}.
\end{equation} 
In total, this gives us four equations, one for every combination of $n,k=A,B$. Differentiating Eq. \ref{kir2b} and substituting in Eq. \ref{eq:kir4} leads to
\begin{widetext}
\begin{equation}
\label{eq:der0}
\frac{ \partial{J_n}} {\partial{\Phi_k}} = I_{0} \left( \cos \gamma_{n-} \cos \gamma_{n+} - \sin \gamma_{n-} \sin \gamma_{n+} \tan \gamma_{n-} \tan \gamma_{n+} \right)\frac{\partial \gamma_{n-}}{\partial \Phi_{k}}.
\end{equation} 
\end{widetext} 
Finally we are left with obtaining $\frac{\partial \gamma_{n-}}{\partial \Phi_{k}}$, which will depend on whether $n$ and $k$ are the same or different. Hence differentiating Eq.~\ref{constraint1}, and taking $n=k$ gives
\begin{equation} \label{eq:der2}
\frac{\partial \gamma_{n-}}{\partial \Phi_{n}} = \frac{\pi}{\Phi_{0}}\left(1 - M_{n}\frac{ \partial{J_n}} {\partial{\Phi_n}}  - M_{nk}\frac{ \partial{J_k}} {\partial{\Phi_n}}\right),
\end{equation} 
and when $n\neq k$
\begin{equation}
\label{eq:der1}
\frac{\partial \gamma_{n-}}{\partial \Phi_{k}} = \frac{\pi}{\Phi_{0}} \left( - M_{n}\frac{ \partial{J_n}} {\partial{\Phi_k}}  - M_{nk}\frac{ \partial{J_k}} {\partial{\Phi_k}} \right).
\end{equation} 
Substituting the results of Eqs. \ref{eq:der1} and \ref{eq:der2} into \ref{eq:der0} leads to

%
%
%

\begin{widetext}
\begin{equation}
\label{eq:djdp1}
\frac{ \partial{J_n}} {\partial{\Phi_n}} = \frac{1}{2L^{J}_{n}} \frac{1-\tan^{2}\gamma_{n+} \tan^{2}\gamma_{n-}}{1 + \frac{M_n}{2L^{J}_{n}} \left[1 - \tan^{2}\gamma_{n+} \tan^{2}\gamma_{n-} \right] } \left( 1 - M_{nk} \frac{\partial{J_k}}{\partial{\Phi_n}} \right)  
\end{equation} 
\begin{equation}
\label{eq:djdp2}
\frac{ \partial{J_k}} {\partial{\Phi_n}} = \frac{-1}{2L^{J}_{k}} \frac{1-\tan^{2}\gamma_{k+} \tan^{2}\gamma_{k-}}{1 + \frac{M_k}{2L^{J}_{k}} \left[ 1 - \tan^{2}\gamma_{k+} \tan^{2}\gamma_{k-} \right] } M_{nk} \frac{\partial{J_n}}{\partial{\Phi_n}} 
\end{equation}
\end{widetext}
where we have taken 
\begin{equation}
\label{eq:djdp3}
L^{J}_{n}=\frac{\Phi_{0}}{2 \pi I_{0}  \cos \gamma_{n-} \cos \gamma_{n+}}
\end{equation} 
to be the Josephson inductance. We easily observe that in the case of one coupler, $ M_{n,k}=0$ in Eq. \ref{eq:djdp1} and Eq. \ref{eq:djdp2} does not come into play. This reduces the result to the one obtained in the original paper \cite{Plourde2004}. Hence, for a given geometry, applied fluxes $\Phi_{A}, \Phi_{B}$ and bias currents $I_{A}, I_{B}$, we now have a means of calculating the Hamiltonian from Eq. \ref{TotalH}.

\section{Coupling Results}
\label{CouplingResults}

In our numerical calculations, we take the qubits' persistent current values to be $I_{p}=0.46 \mu$A and the critical currents of the Josephson junctions that are used in DC-SQUIDs as $0.11 \mu$A --- both in the experimentally achievable ranges. In order to answer whether all couplings can be turned off, or whether we can selectively make certain interactions strong, while keeping others weak, we need to understand what happens to all the coupling terms for all possible input parameters. To do this, we scan both of the applied fluxes $\Phi_A$ and $\Phi_{B}$ between $-\Phi_0 /2$ and $ \Phi_0 /2$ and the bias currents between $0$ and values approaching the critical currents of the couplers (which depend on the threaded fluxes). The inductance matrix is calculated for each geometry and used to obtain values for all possible $K_{ij}$. For example, in the L-shaped geometry with the thin arm width of $w=2\mu$m and the arm length of $a=50 \mu$m, its value is

\begin{widetext} 
\begin{equation}
 \bordermatrix{
  & Q1 & Q2 & Q3 & A & B \cr
Q1 & 171.5483 &     -0.4689 &     -0.1659 &    73.9081 &     -0.8751   \cr
Q2 & -0.4689 &   171.5483 &    -0.4600 &   75.6132 &    75.5106   \cr
Q3 & -0.1659 &    -0.4600 &   171.5483 &     -0.8751 &    73.9081   \cr
A & 73.9081 &   75.6132 &     -0.8751 &   457.8219 &   91.8483   \cr
B & -0.8751 &    75.5106 &    73.9081 &   91.8483 &   457.8219   \cr
} \ \text{pH},
    \label{eq:IndMatrix250}
\end{equation}
\end{widetext}
where the columns (and rows) correspond to the first, second, third qubits, coupler A, and coupler B respectively.

Fig.~\ref{fig:ksdata} shows the coupling strength of the L-shape geometry, where the arm width ``$w$'' is $2\mu$m and the arm length ``$a$'' is $100\mu$m. The horizontal axis corresponds to the values of coupling energies $K_{12}$, the vertical to $K_{23}$ and finally the ``out-of-page'' direction to the crosstalk term $K_{13}$. The plot was obtained by varying the fluxes and biasing currents through both couplers. Thousands of data points were calculated, which were then used to obtain a surface that spans different combinations of the three coupling energies. The process was then repeated for varying arm lengths and arm widths of the coupling DC-SQUIDs. From the results one can note that the crosstalk term $K_{13}$ can be treated as a single-valued function of $K_{12}$ and $K_{23}$. We will use this fact in the next section.

\begin{figure}[th]
\includegraphics[width=3.2in]{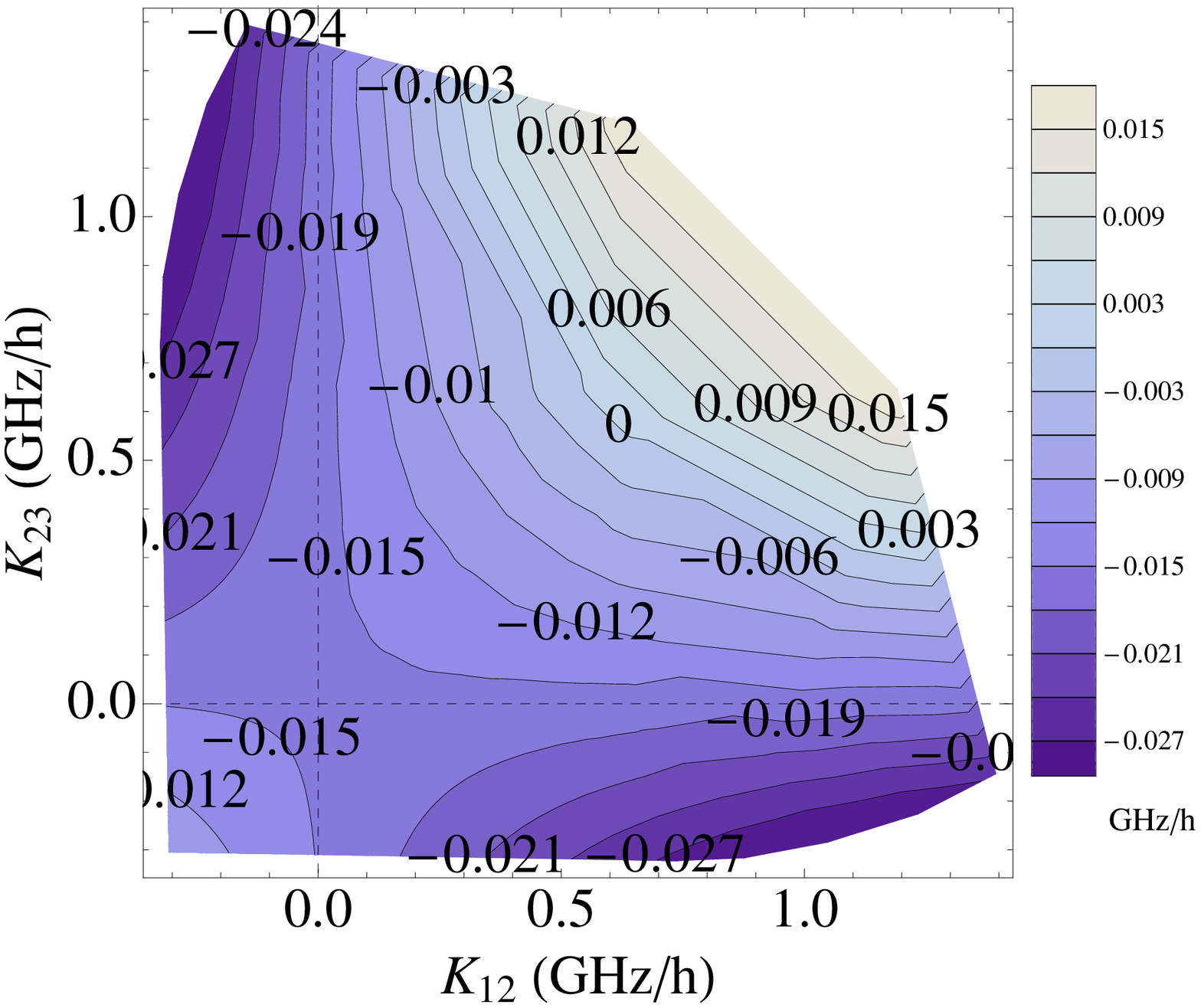}
\caption{(Color online). Interaction energies $K_{12}$ (horizontal axis), $K_{23}$
  (vertical axis) and $K_{13}$ (out-of-page direction) for an L-shape
  geometry with couplers' arm width of $w=2 \mu$m and arm length
  $a=100 \mu$m. In the calculations we
  take $I_1=I_2=I_3=0.46 \mu$A and $I_0=0.11 \mu$A.
  }
\label{fig:ksdata}
\end{figure}

The first question we wish to address is whether we can keep the coupling between two nearest-neighbour qubits (say 1 and 2) high, the other nearest-neighbour coupling (say 2 and 3) turned off, while eliminating the crosstalk (interaction between qubits 1 and 3). We concentrate on this scenario as we envision doing gates between nearest-neighbors (here qubits 1 and 2), and would like the qubits that are not involved in the gate (qubit 3) to be disturbed as little as possible. In practice, we will find that it is not possible to completely turn off the interactions with the third qubit, and hence all the values of $K$ will need to be varied, but during a gate between qubits 1 and 2 the interaction $K_{12}$ will dominate. We can get the answer to this first question by studying plots like the one presented in Fig.~\ref{fig:ksdata}. We first fix $K_{23}$ (vertical axis) at zero, and then traverse in the horizontal direction along the dashed line. Figs.~\ref{fig:line_shaped_w2} and \ref{fig:lshaped_w2} show cross-sections for the line-shaped and L-shaped geometries, with the arm width $w=2\mu$m. 
In the line-shaped geometry scenario (Fig.~\ref{fig:line_shaped_w2}), with shortest
considered arm length $a=25\mu$m and $K_{23}$ tuned to approximately
0, we find the crosstalk $K_{13}$ to be $47$MHz. The
crosstalk only varies slightly as we increase the coupling $K_{12}$.
In the L-shape geometry (Fig.~\ref{fig:lshaped_w2}), the
situation is less favorable, as expected, since qubits 1 and 3 are
closer together and their direct mutual inductance is larger.
Furthermore, they interact with the coupler that is further away from
them more than is the case in the line-shaped geometry. From Fig.~\ref{fig:lshaped_w2} we
see that the best we can do given short couplers with $a=25 \mu$m is
$K_{13} \approx 160$MHz.
For concreteness, if one considers a ``high'' value of coupling 1-2 to be $\approx 1.0$GHz, these results lead to the unwanted crosstalk of 5--16\% percent when the qubits are closest together. However, as expected and shown in Figs. \ref{fig:lshaped_w2} and \ref{fig:line_shaped_w2}, we can reduce the unwanted crosstalk term $K_{13}$ by simply increasing the arm length of the coupling DC-SQUIDs --- for example in a case of $a=300 \mu$m, for $K_{12} > 1.0$GHz, the magnitude of $K_{13}$ is smaller than $2$MHz in all the geometries. We do stress however that one cannot keep increasing the arm length indefinitely, as the process also decreases the maximum achievable energy that can be mediated between nearest-neighbours qubits. We have further found that increasing the arm width of the couplers to $w=48\mu$m made no qualitative difference in the results above, however for the same set of parameters the magnitudes of the interaction energies $K_{ij}$ were smaller. This is due to the fact that the mutual inductance elements between these wider coupling DC-SQUIDs and qubits are smaller than in cases of the narrow couplers. The discussion above leads us to conclude that minimizing the crosstalk coupling (term $K_{13}$) can be done to a great degree, even when fairly small DC-SQUIDs are used and more importantly in both of the studied geometries. 

\begin{figure}[ht]
\includegraphics[width=3.2in]{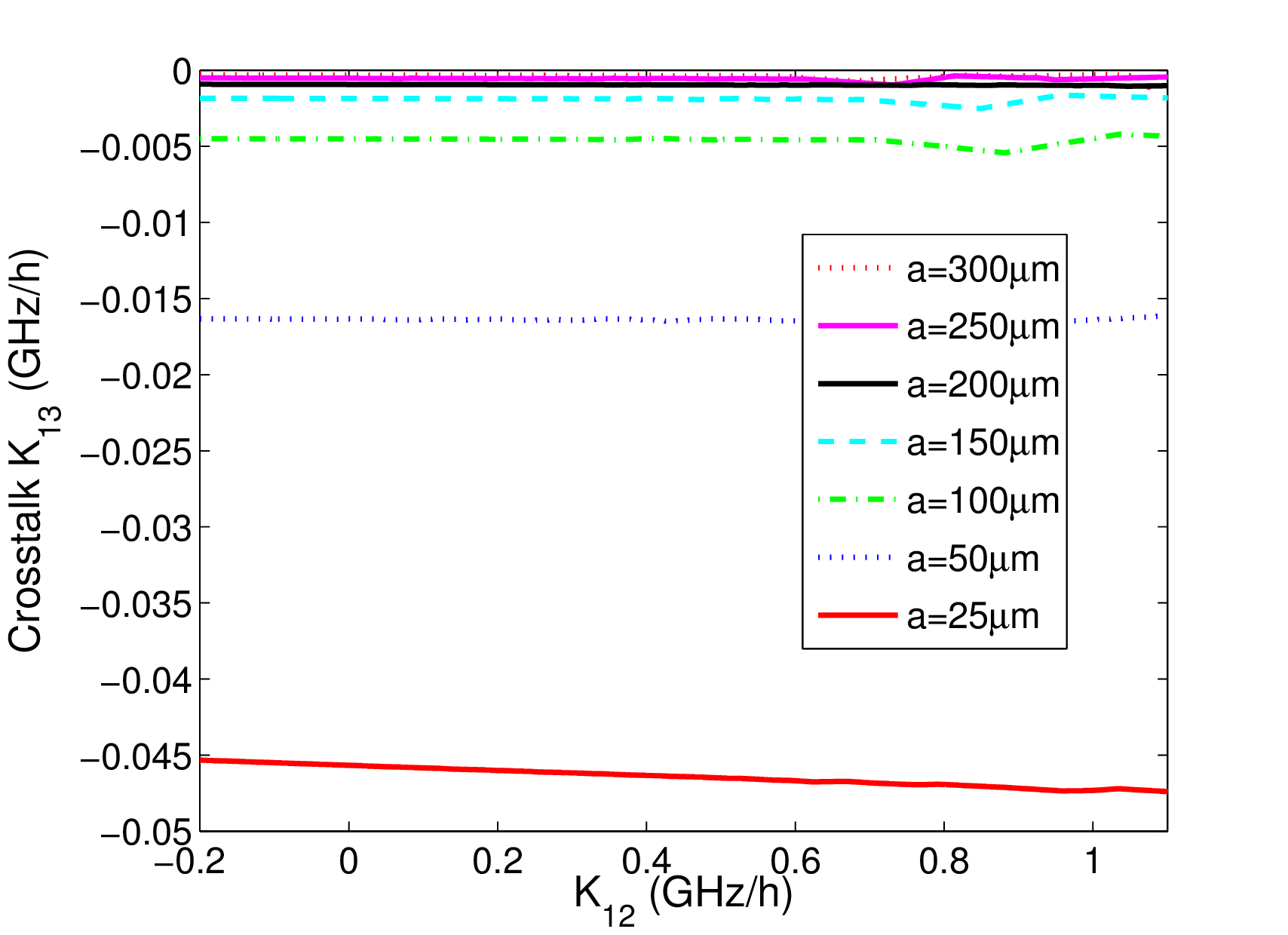}
\caption{(Color online). The size of the crosstalk interaction $K_{13}$ as a function
  of $K_{12}$ in the line-shaped geometry scenario where the coupling
  $K_{23}$ has been taken to be zero. In the calculations we
  take $I_1=I_2=I_3=0.46 \mu$A and $I_0=0.11 \mu$A. }
\label{fig:line_shaped_w2}
\end{figure}

\begin{figure}[ht]
\includegraphics[width=3.2in]{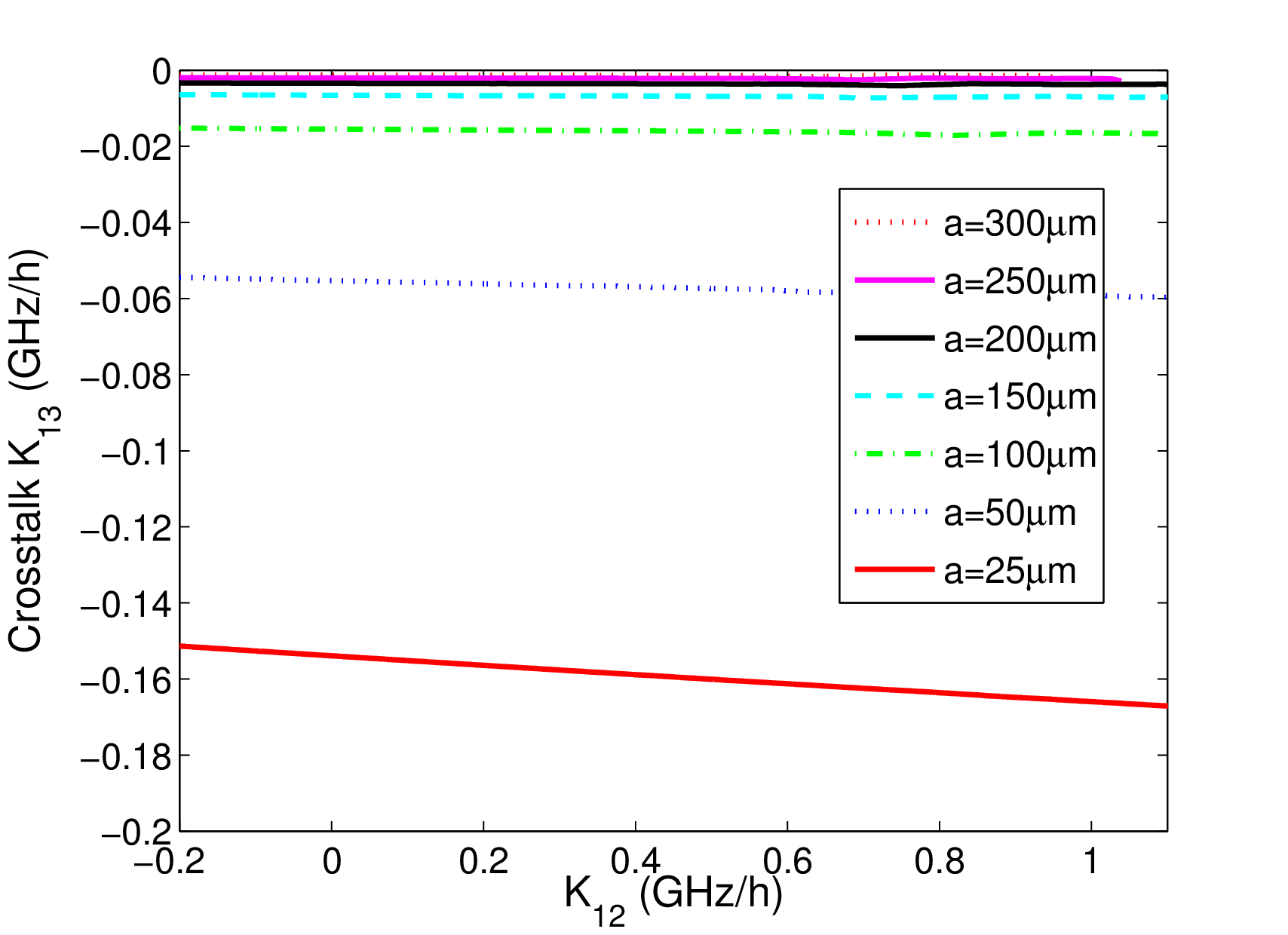}
\caption{(Color online). The size of the crosstalk interaction $K_{13}$ as a function of $K_{12}$ in the L-shaped geometry scenario where the coupling $K_{23}$ has been taken to be zero. In the calculations we take $I_1=I_2=I_3=0.46 \mu$A and $I_0=0.11 \mu$A.}
\label{fig:lshaped_w2}
\end{figure}


The next question we wish to answer is whether the interaction between all pairs of qubits can be turned off completely. This may be useful when one wants to perform single gate operations on any (or all) of the qubits without affecting the others. To determine this, we once again turn to Fig.~\ref{fig:ksdata} (and its analogues for other geometries and values of arm lengths and widths --- not explicitly shown). We can now look at the crossing of the two dashed lines, which corresponds to $K_{12}=K_{23}=0$, and figure out the value of the crosstalk ($K_{13}$). We plot the results in Fig.~\ref{fig:K13AtZero} for both geometries and two different arm widths.
\begin{figure}[!h]
    \centering
    \includegraphics[width=7cm]{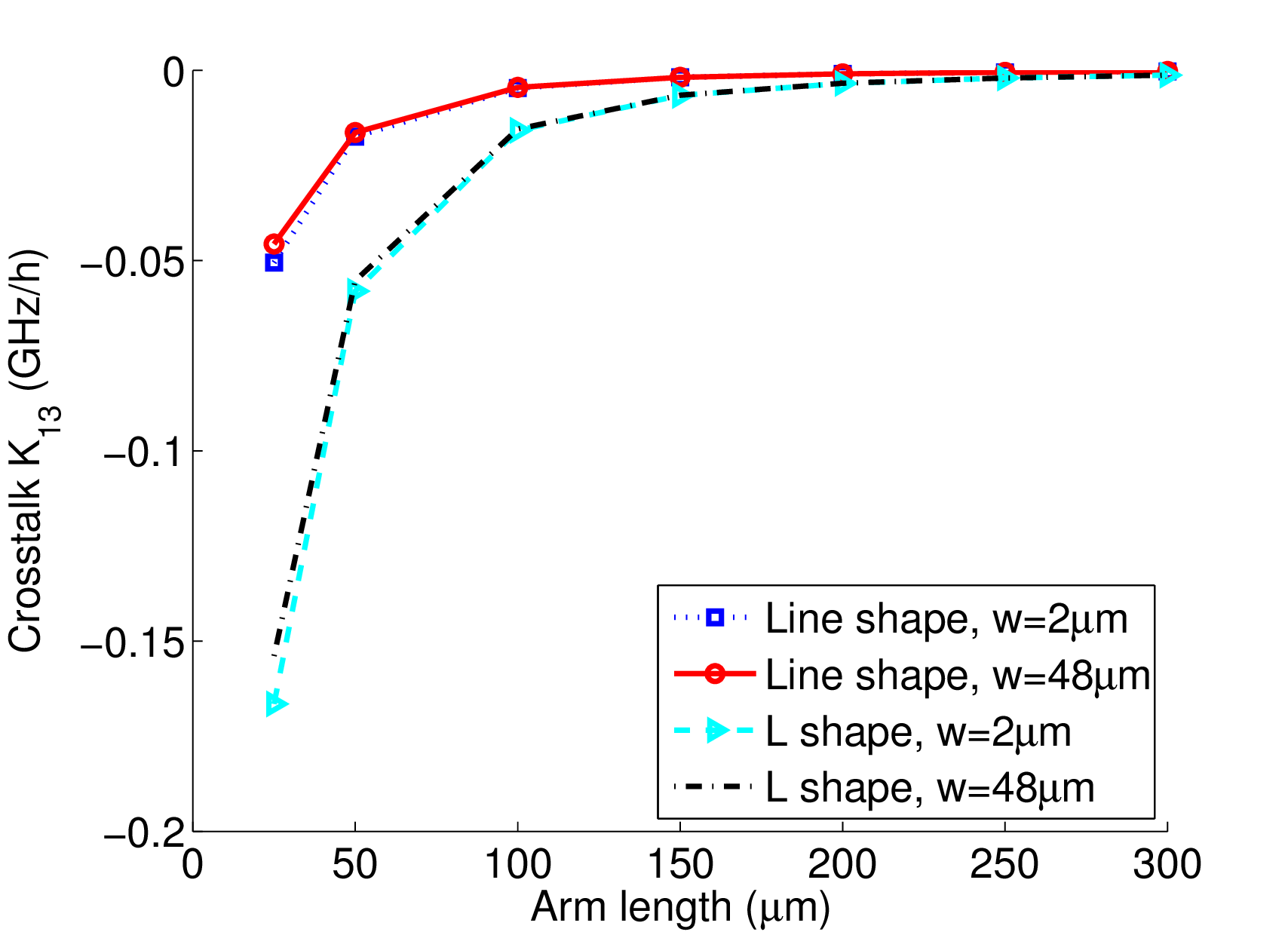}
    \caption{(Color online). A plot consisting of values for the crosstalk $K_{13}$, while keeping $K_{12}=K_{23}\approx 0$. The horizontal axis represents different arm lengths of the coupling DC-SQUIDs. We find that the magnitude of $K_{13}$ is substantially smaller in the case of the straight-line geometry, but even in the L-shaped scenario dies off quickly. In the calculations we
  take $I_1=I_2=I_3=0.46 \mu$A, $I_0=0.11 \mu$A.}
    \label{fig:K13AtZero}
\end{figure}
From the data it is clear that there is no complete off-state. The straight-line geometry is much less (over 3 times in a case of shortest arm length we studied) susceptible to the unwanted crosstalk $K_{13}$. However the crosstalk energy dies off quickly, and in a case of arm length $a>200 \mu$m, its magnitude stays below $5 {\rm MHz}$. Given that the typical energies of the qubits are of the order of a few GHz, a crosstalk strength of roughly three orders of magnitude smaller may not be very damaging. We also find that the width of the couplers has little effect on the on-off state. 
We further note that the crosstalk and off-state fidelity could in principle be improved by
additional superconducting screening elements \cite{Harris10}, which are not
examined in our work. Others have also shown that there is a distinct crosstalk
mechanism related to the breakdown of the adiabatic approximation,
discussed, e.g., in \cite{hutter2006tcq} and exemplified in \cite{Pinto10}. 
We chose, however, in this work to explore control that
tolerates the crosstalk at hand. 
\section{Logical Operations}
\label{QubitControl}
A crucial component of any quantum computer is its ability to perform
logical operations. These operations can be constructed from only
single-qubit rotations and (at least) one two-qubit gate (such as a
CNOT) \cite{Nielsen00}. In \cite{Plourde2004} it was shown how a two-qubit gate can be performed using results from \cite{Zhang03}. Here, in the three-qubit case, we concentrate on a different approach. We use numerical optimization via the GRAPE algorithm \cite{Khaneja05} to find a collection of pulses that maximize the gate fidelity $F_{g}$, with
\begin{equation}
	\begin{split}\label{eq:fid}	
        F_g	= \text{Tr} \{ U \dg U_N \}^{2}
	\end{split}
\end{equation} 
where $U$ is the desired gate, and $U_{N}$ a calculated numerical approximation. GRAPE works by dividing the gate time into time-slices (often called time pixels) of constant pulse amplitude. The amplitudes at any time pixel are found by calculating the gradient of the fidelity function with respect to some well defined controls. Since the gradient points in the direction of increasing $F_{g}$, over many iterations, one can converge to a pulse sequence that is (arbitrarily) close to the intended gate.  
We look at two gates, the first, a ``natural'' candidate for this type of system is the $\sqrt{\text{iSWAP}}$, and the other a CNOT gate. We stress that a CNOT gate is often used to describe the active error correction procedures in the topologically motivated schemes from \cite{Kitaev1995,Raussendorf2007b,Fowler2009}.
\begin{figure}[ht]
\includegraphics[width=3.4in]{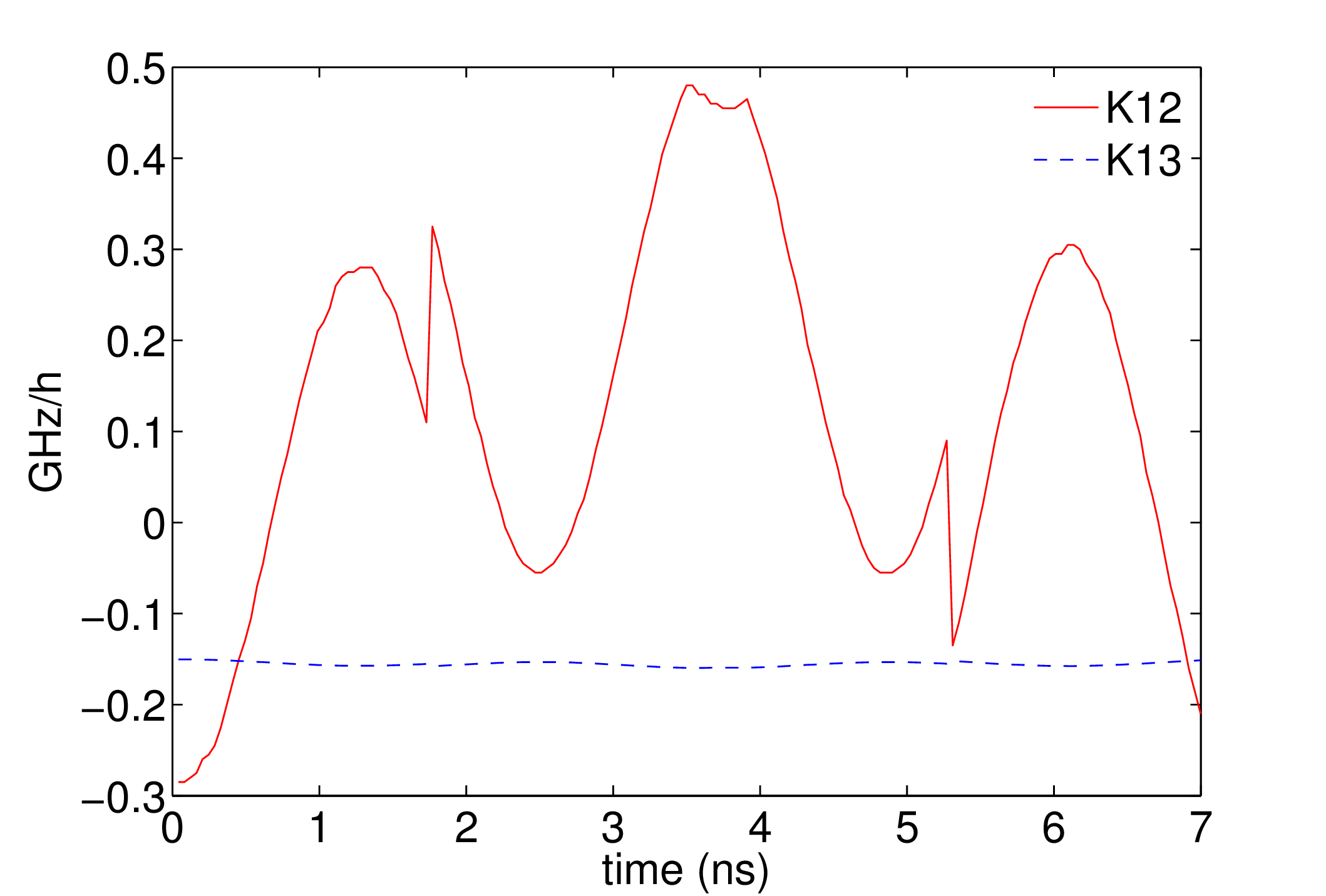}
\caption{(Color online). A pulse sequence for a $\sqrt{\text{iSWAP}}$ gate between qubits 1 and 2. Since the system coupling is ``natural'' for this gate, no single-qubit rotations are needed. The pulses lead to the desired gate with 99.9\% fidelity. In the calculations we take $\Delta_1/h=5.0 {\text{ GHz}}$, $\Delta_2/h=5.4 {\text{ GHz}}$ and $\Delta_3/h=5.8 {\text{ GHz}}$.}
\label{fig:iswap}
\end{figure}

We start with the system Hamiltonian shown in Eqs.~\ref{QubitH} and \ref{TotalH}, and note that each $\epsilon_j$ is time dependent and can be written as
\begin{equation}
 \epsilon_j(t)= 4 \epsilon^{x}_{j}(t) \cos(\omega_j t) + 4 \epsilon^y_j(t) \sin(\omega_j t).    
    \label{eq:Epsilons}
\end{equation}
Having both $\epsilon^x_j(t)$ and $\epsilon^y_j(t)$ present, will later give us the $\sigma_x$ and $\sigma_y$ quadratures. We use both of these when creating CNOT pulses. We can now switch basis, to the eigenbasis of $\sigma_x$, namely $\{\ket{+}, \ket{-}\}$ and furthermore go into a rotating frame defined by $U= U_{1} \otimes U_2 \otimes U_3$ with $U_j = \exp (-i \frac{\Delta_j}{2} \sigma^j_z t )$. We then assume that the qubits can be driven at their respective resonances (meaning $\Delta_j=\omega_j$) and neglect the fast oscillating terms (do a RWA). All this leads to an effective Hamiltonian in the rotating frame and $\{\ket{+}, \ket{-} \}$ basis

\begin{equation}
\begin{split}
    H_{e}=& \sum_{j} \epsilon^{x}_j (t)\sigma^j_x -  \epsilon^{y}_j  (t)\sigma^j_y \\
        & -K_{12}(t) e^{-i(\Delta_{1} - \Delta_{2})t} \left( \ket{100} \bra{010} + \ket{101} \bra{011} \right) + h.c. \\
        & -K_{23}(t) e^{-i(\Delta_{2} - \Delta_{3})t} \left( \ket{001} \bra{010} + \ket{110} \bra{101} \right) + h.c. \\
        & -K_{13}(t) e^{-i(\Delta_{1} - \Delta_{3})t} \left( \ket{100} \bra{001} + \ket{110} \bra{011} \right) + h.c..
\end{split}
    \label{eq:Heff}
\end{equation}
We explicitly make the $K_{ij}$ terms time dependent to stress that they will vary.
For demonstration purposes, in the case of $\sqrt{\text{iSWAP}}$, we chose the qubit energies to be $\Delta_1/h=5.0 {\text{ GHz}}$, $\Delta_2/h=5.4 {\text{ GHz}}$ and $\Delta_3/h=5.8 {\text{ GHz}}$, and when creating the pulses for the CNOT gate we take all qubits to be on resonance --- meaning $\Delta_1/h=\Delta_2/h=\Delta_3/h=5.0 {\text{ GHz}}$.

We constrain GRAPE by only allowing values of $K_{ij}$ which can be achieved using physical controls (as described in section \ref{CouplingResults}). We can think of this as staying on a surface spanned by the triplet $(K_{12}, K_{23}, K_{13})$. From Section \ref{CouplingResults}, we note that $K_{13}$ can be treated as a single-valued function of $K_{12}$ and $K_{23}$. This slightly simplifies our procedure of finding the gradient of $F_{g}$, since we are only dealing with two independent variables that account for all the interactions in the problem. We further allow GRAPE to vary single-qubit controls (which effectively means $\epsilon^x_j(t)$ and $\epsilon^y_j(t)$ from Eq.~\ref{eq:Heff}).

The resulting pulses for the $\sqrt{\text{iSWAP}}$ and CNOT gates are presented in Figs.~\ref{fig:iswap} and \ref{fig:cnot} respectively. Only the non-zero controls are explicitly shown. We study the case of the L-shaped geometry with the arm length of $25\mu$m where the unwanted crosstalk coupling $K_{13}$ is the largest (and as can be seen from the plots, never zero). 

In the case of $\sqrt{\text{iSWAP}}$, it is clear that only coupling 1-2 has to be varied, and no single-qubit rotations are needed on any of the qubits. Since we have chosen the qubits to be off-resonance with each other, in the rotating frame the terms $K_{23}$ as well as the crosstalk $K_{13}$ oscillate at frequencies  $(\Delta_2 - \Delta_3)$ and $(\Delta_1 - \Delta_3)$ respectively. Because they are not too large and average out to zero over the length of the pulse, little correction is needed in the generated controls.

The situation is more complicated in the case of the CNOT gate, partly because it is a ``non-natural'' gate for the Hamiltonian we have, but also due to the fact that here we chose to keep all qubits on-resonance with one another. The resulting pulses require single-qubit controls and in particular both $\sigma_x$ and $\sigma_y$ quadratures on qubit 3 need to be varied. In contrast to the off-resonant case, to achieve high fidelities, qubit 3 must be repeatedly flipped and rotated to counteract the crosstalk term.

In both cases, the pulses found perform with a fidelity of $99.9\%$. That number however can be often increased (in which case the calculation can take substantially longer). The pulses in general are not unique and can depend on the initial guesses which GRAPE then modifies. 

\begin{figure}[ht]
\includegraphics[width=3.4in]{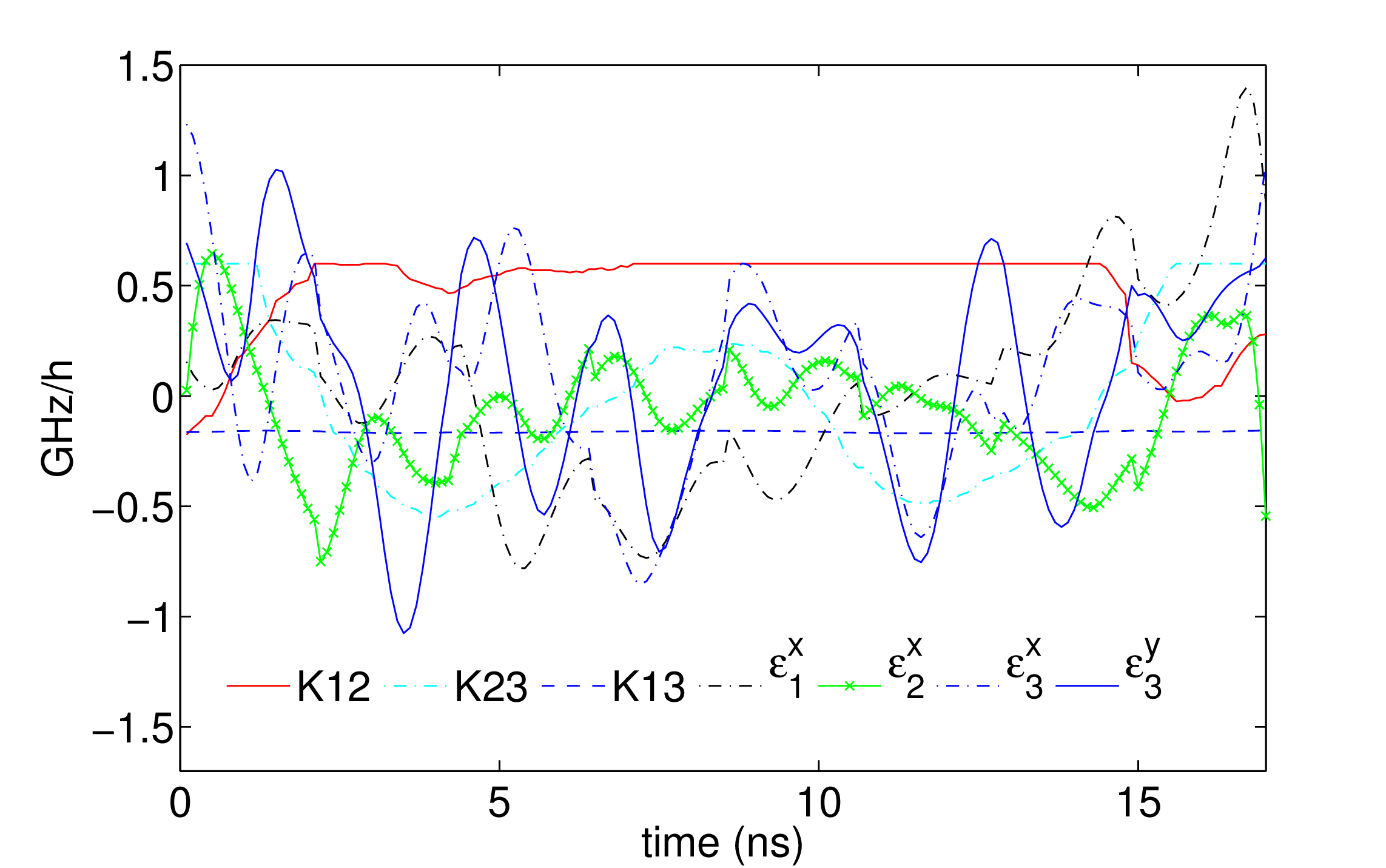}
\caption{(Color online). A pulse sequence for a CNOT gate between qubits 1 and 2. Single-qubit rotations are applied to all three qubits. For timely convergence of the GRAPE algorithm, we modulate both $\sigma_x$ and $\sigma_y$ quadratures on the third qubit. Furthermore, we allow all three interaction energies to vary. The pulses lead to the desired gate with 99.9\% fidelity. In the calculations we take $\Delta_1/h=\Delta_2/h=\Delta_3/h=5.0 {\text{ GHz}}$.}\label{fig:cnot}. 
\end{figure}

Finally, let us note that we have shown how one has to vary the energies in the system to obtain the desired pulses. Alternatively one could concentrate on working with values of physical parameters (currents, fluxes) which might be more useful when trying to devise an experimental demonstration of this system.

\section{Conclusion}
\label{Conclusion}

In this paper, we discussed a system which may provide a candidate for
a  many-qubit processor built out of superconducting circuits. In
particular we extended the coupling mechanism first presented in
\cite{Plourde2004} to a three-qubit, two-coupler scenario. We
studied multiple geometries of the coupling DC-SQUIDs; a line-shape,
and an L-shape, with couplers' arm lengths between $25\mu$m and $300
\mu$m, and arm widths of $2$ and $48 \mu$m. We looked at how much
unwanted crosstalk ($K_{13}$) there is while the nearest-neighbour
interaction ($K_{12}$ or $K_{23}$) is high ($\approx 1$GHz).
 With our experimentally achievable physical parameters, when the arm lengths are shortest, we found the maximum crosstalks of 
47MHz and 160MHz for the line-shape and L-shape geometries
 respectively. We also characterized the crosstalk in the off-state
 and found that for arm lengths greater than $200\mu$m it was smaller
 than 5MHz, which is three orders of magnitude smaller than typical
 qubit energies.
 Finally, we showed how one can use optimal control methods (GRAPE) to perform logical operations in the presence of crosstalk (while accounting for the constraints on all the coupling energies), and in particular we showed potential pulse sequences for the $\sqrt{\text{iSWAP}}$ and the CNOT gates. 

\section{Acknowledgements}
PG, FM and FKW were supported by NSERC through the
discovery grants. This research was also funded by the Office of the Director of National Intelligence (ODNI), Intelligence Advanced Research Projects Activity (IARPA), through the Army Research Office. All statements of fact, opinion or conclusions contained herein are those of the authors and should not be construed as representing the official views or policies of IARPA, the ODNI, or the U.S. Government. AGF acknowledges support from the Australian Research Council Centre
of Excellence for Quantum Computation and Communication Technology
(Project number CE110001027).

\bibliographystyle{./mybibtexstyle}
\bibliography{../reference1}

\end{document}